\documentclass[12pt]{article}

\usepackage{amsmath,epsfig,epsf,psfrag,natbib}
\usepackage{amssymb,url}
\usepackage[latin1]{inputenc}
\usepackage{amsbsy}
\usepackage{lscape}
\usepackage{color,latexsym,amsfonts,amsthm,amscd,graphicx}
\usepackage{mathrsfs}
\usepackage{bbm}
\usepackage{array,multirow}

\newcommand{\be}{\begin{equation}}
\newcommand{\ee}{\end{equation}}
\newtheorem{theorem}{Theorem}

\newtheorem{example}{Example}
\newtheorem{remark}{Remark}

\newcommand{\ba}{\begin{eqnarray}}
\newcommand{\ea}{\end{eqnarray}}
\newcommand{\bas}{\begin{eqnarray*}}
\newcommand{\eas}{\end{eqnarray*}}
\newcommand{\ben}{\begin{enumerate}}
\newcommand{\een}{\end{enumerate}}

\def\T{{ \mathrm{\scriptscriptstyle \top} }}

\newcommand{\bx}{ { \bf  x}}
\newcommand{\by}{ { \bf  y}}
\newcommand{\bq}{ { \bf  q}}

\newcommand{\btheta}{\mbox{\boldmath $\theta$}}
\newcommand{\blambda}{\mbox{\boldmath $\lambda$}}
\newcommand{\ind}{\mathbbm{1}}

\include{commands_jz}

\begin{document}
\title{Permutation tests under a rotating sampling plan with clustered data}
\author{Jiahua Chen$^*$, Yukun Liu$^\dag$, and James Zidek$^*$\\
$^*$University of British Columbia and $^\dag$East China Normal University}
\date{\today}
\maketitle

\begin{abstract}
%{\bf Summary}

%\medskip
{
Consider a population consisting of clusters of sampling units,
evolving temporally, spatially, or according to other dynamics.
We wish to monitor the evolution of its means, medians, or other parameters.
For administrative convenience and informativeness, 
clustered data are often collected via a rotating plan.
Under rotating plans, the observations in the 
same clusters are correlated, and observations on 
the same unit collected on different occasions are also correlated.
Ignoring this correlation structure
may lead to invalid inference procedures.
Accommodating cluster structure in parametric models
is difficult or will have a high level of misspecification risk.
In this paper, we explore exchangeability in clustered data 
collected via a rotating sampling plan to develop
a permutation scheme for testing various hypotheses of interest.
We also introduce a semiparametric density ratio model
to facilitate the multiple population structure in rotating sampling plans.
The combination ensures the validity of the inference methods
while extracting maximum information from the sampling plan.
A simulation study indicates that the proposed tests firmly
control the type I error whether or not the data are clustered. 
The use of the density ratio model improves the power of the tests.
}
\end{abstract}


\section{Introduction}
In many problems, including the application (see Section~\ref{sec:motivator})
that motivated this paper, clustered data are collected via a rotating 
sampling plan.
Such a plan provides administrative convenience and informativeness \citep{nijman1991efficiency}
and is a common tool in sampling practice 
\citep{visagie2019measuring}.
Under rotating plans, the observations in the 
same clusters are correlated, and observations on 
the same unit collected on different occasions are also correlated.
Ignoring this correlation structure
may lead to invalid inference procedures.
One must develop specific statistical theory and 
methods to correctly and effectively draw information from the data.

However, in the above context, population parameters such as 
percentiles of practical interest e.g., in structural engineering, 
lie outside the domain of  applicability of classical statistical methods. 
Clustered data prove particularly challenging, and
ignoring their structure leads to inflated type I 
errors \citep{verrill2015simulation,datta2008signed}.  The result will be 
unacceptably high frequencies of false rejections, i.e., false alarms.
It is difficult to build defensible and easy-to-use parametric models that  
accommodate complex cluster structures.
Some recent results such as those of \citet{chen2016monitoring}
are limited to handling independent cross-sectional
samples from multiple populations with or without clusters.
The goal of this paper is to develop a method
capable of handling the longitudinal random effects implied by the 
rotating sampling plan.

We develop a method for testing a null hypothesis $ H_0 $ concerning 
population parameters against an alternative hypothesis $ H_a $. 
However, we diverge from the conventional Neyman--Pearson (NP) approach of 
developing a test statistic and then
finding its rejection region and thereby choosing the appropriate action 
in a decision-theoretical framework, selecting either the null or alternative hypothesis. 
The complexity of the inferential problems forces us to seek an alternative approach.  
We begin by borrowing from the Fisherian alternative to the NP approach: 
significance testing \citep{johnstone1987tests}. 
Thus, we first define a test statistic to partner with the null hypothesis 
but with the alternative in mind.  
Given the data, we then generate a p-value for that statistic. 
We depart from significance testing by using our p-value merely 
as a way to find the rejection region for the NP test, 
since an analytical determination of that region is generally 
not feasible except possibly by invoking asymptotic theory.  

Finding the p-value presents its own challenges, 
if we follow the conventional route of specifying a sampling distribution 
and then its cousin, the likelihood function, and so on.  
So we again borrow from Fisher, who argued in favor of 
separating the sampling model from the inferential model.  
Thus, his so-called exact test for contingency tables was born. 
Our approach uses a permutation scheme for hypothesis testing that
follows from the assumption that the sequences of samples  are   
exchangeable over time or space in data collected  via a rotating sampling plan.  
Surprisingly, this modest assumption enables us to empirically  
compare the performance of each member of 
a rich class of possible test statistics. 
Moreover, the approach proves to make modest computational demands.

In fact, the permutation is a general approach that plays an active role in
modern statistical practice \citep{pesarin2010permutation, hemerik2018, hemerik2019permutation}.
To provide the stochastic foundation for the test, we use the semiparametric 
density ratio model or DRM
\citep{anderson1979multivariate,qin1997goodness}
to accommodate the multiple population structure.
Then we use the empirical likelihood \citep[EL;][]{owen2001empirical} to 
construct nonparametric test statistics.
A simulation study indicates that the  tests  proposed in this paper firmly
control the type I errors whether or not the data are clustered. 
The use of the DRM improves the power of the tests.
An investigation of the influence of 
the choice of basis function in the DRM
suggests that the efficiency gain is widely observed.

The paper is organized as follows.  In the next section, we describe
the rotating sampling plan and its implied random effects. 
We then reveal the structural symmetry 
in multiple populations and their clusters. 
Section \ref{permutation} gives details of the proposed generic permutation test. 
Section \ref{statistics} proposes a number of test statistics to be used
for the permutation tests, including the straightforward and classical
t-statistic and Wilcoxon rank-sum statistic. Their type I errors
become well controlled when their p-values are computed through
a permutation scheme, rather than through asymptotic theory developed
for independent and identically distributed (IID) observations.
The simulation experiments in Section~\ref{simulation} 
show that the classical t-test and Wilcoxon rank-sum test have inflated type I
errors if the p-values are computed by the asymptotic theory developed
for IID observations.
In contrast, the permutation tests based on various test statistics,
including the t-statistics and Wilcoxon statistics, are shown to have well-controlled type I errors.
A combination of the DRM and the EL
leads to tests for percentiles that have much improved powers. 
This advantage obtains even when
the basis function underlying the DRM is mildly mis-specified.
The paper concludes with a discussion and an appendix
that gives technical details for the numerical methods
employed in the simulation experiments.

\section{Motivating application}\label{sec:motivator}
Lumber is manufactured  from the trees of a forest.  
The trees are cut down and trimmed to get logs that are transported to 
mills where they are sawn in an optimal way to get pieces of lumber. 
These pieces are classified into grades, primarily according to their strength 
for engineering applications. 
Each grade has a published design value (DV)  
for each type of strength, notably under stretching (ultimate tensile strength or UTS), 
compression, or bending (modulus of rupture or MOR).  
Its stiffness (modulus of elasticity or MOE), which is related to all these other characteristics, 
is, unlike MOR, not measured by destructive testing.  

The DV is a specified quantile of the strength 
distribution, commonly a median or the fifth percentile. 
Thus, the grade of a piece of lumber for engineering applications
depends on its intended use.  
The top grade is both  strong and expensive.
The development of the modern grading system 
has been a triumph of structural engineering 
since it has standardized lumber properties.  
Thus, wood, a heterogeneous material unlike say aluminum, 
can be used with the assurance that the lumber made from it 
has a low probability of failure when used for its intended purpose.

Given the changing climate and its consequences such as changes in the way trees 
grow, forest fires, and insect infestations, the quality of wood  
may vary across regions and over time. 
Along with this has come changes in processing techniques.  
Thus, concern has arisen about a possible decline in the strength of lumber.  
This has led to the need to estimate and hence monitor DVs over time. 
The first such long-term monitoring program was established in 1994 
in the southeastern United States. 
Cross-sectional samples were taken annually using a stratified-by-region sampling plan. 
The number of mills in each region was determined, and
the primary sampling units (PSUs) within a region were chosen by simple random sampling. 
One or two bundles, i.e., secondary sampling units (SSUs), of about $ 300 $ pieces 
each were selected. From each, a ``lot'' with ten pieces was chosen in a prescribed way 
and its MOE was measured. 
Since the MOE does not involve destructive testing, 
the sampling plan was relatively inexpensive. 
Moreover, the MOE was predictive of the other strengths. 
The subsequent analyses used paired t-tests, since the MOE has 
an approximately normal distribution.  

Canada also established a long-term monitoring program. 
Planning for a pilot program, also measuring the MOE,  
began in 2005; a preliminary analysis  showed a substantial 
variation between mills, within mills, and between lots.  
The goal at the time was to measure temporal trends in the MOE, and so
a rotating panel design was selected for a specified grade of lumber,
with a six-year rotation.
This plan limited the mill response burden, 
made a consistent random mill effect over time (six years) plausible,
and refreshed the sample to maintain some degree of cross-sectional validity. 
On the other hand, as for the paired t-test, changes over time could 
with confidence be ascribed to changes in strength rather than merely 
changes in the sample of mills. 
It soon became apparent that the plan established for the MOE  
could be used for monitoring the MOR as well. 

This led to new challenges: the statistical theory 
needed to assess trends in the MOR under a rotating sampling plan did not exist.  
The Forest Products Stochastic Modelling Group (FPSMG), 
based at the University of British Columbia, was therefore established.  
It was co-funded by FPInnovations, a nonprofit industrial research lab, and 
the Natural Sciences and Engineering Research Council of Canada. 
The FPSMG, which has involved engineers and wood scientists at 
FPInnovations working in collaboration with statistical faculty and students, is in its tenth year  
at the time of writing. 
It has made numerous contributions to the theory and practice of 
strength measurement and monitoring for forest products: see for example \citet{zidek2018wood}, \cite{cai2017hypothesis}, \citet{chen2016monitoring},
and \citet{chen2013quantile}.

Meanwhile, for reasons beyond the scope of this paper, 
a separate North American long-term monitoring program has been specified 
in a revision of an ASTM standards document (D1990). 
It assumes a cross-sectional sample once every five years and 
specifies among other things that a Wilcoxin test is to be used to assess 
change in the fifth percentile of the MOR. 
The document ignores both the PSU and SSU cluster effects induced 
by their random effects.   In a companion article to this one, an alternative method 
has been proposed for use in the new ASTM monitoring plan.  
This paper addresses the assessment of trends in strength percentiles 
for rotating panel designs where samples are taken every year. 
Its genesis lies in the need for a method that can handle the cluster effect 
across time and across space.  
 
\section{Rotating sampling plans and their random effects}
\label{setting}
Our research focuses on the analysis of data from rotating sampling plans. 
At its foundation lies a grand population made up of PSUs.
Each such unit is made up of a number of SSUs.
The size of the population is so large that it can be regarded as infinite for practical purposes.
The grand population itself remains stable in terms of 
its PSU composition.
Multiple samples are formed by data collected
from SSUs whose responses evolve over time,
space, or other dimensions. For definiteness and expository simplicity, 
we will assume that the dimension is time.
\vspace*{-2ex}

\subsection{Rotating sampling plan}
In a rotating sampling plan, we first randomly select $ n = m N$ (primary) sampling
units from the population. 
For instance, in year $ 1 $, we may sample $ n = 20 = 4 \times 5 = m \times N$ 
schools as the PSUs.  
We then
collect data from $ r=3 $ students at each of the sampled schools. 
In year $ 2 $ we refresh the sample by 
selecting $ m = 4 $  new schools while retaining $ 16 $ of the original schools. 
Again, $r= 3 $ students are selected from each school.  
This process continues until all $ 20 $ of the original schools 
are replaced, which occurs at year 
$ N  + 1= 6 $. 

Since the populations in applications
are very large, we may consider the sampling at this stage to be done with replacement.
This is well justified in the survey context \citep{RaoShao1992}.
In general, on each new occasion, $ m $ units 
from the rest of the population 
are randomly selected to replace $ m $ units in the original sample. 
Conceptually, this procedure will continue forever. 
After $ N $ or more occasions, all the units in the original sample are
flushed out. 
%The units in the sample are rotated out $m$ at a time.

The adoption of the rotational sampling plan leads to both
longitudinal and cross-sectional clustering structures in the data.
On each occasion, we collect data on only $ r $ ultimate sample
units in the $ k $th sampling unit.
The response values can therefore be denoted
$ \by_{k,i} = (y_{k, i, 1}, \ldots, y_{k, i, r})^\tau $.
For instance, we may collect data from $r=3$ students at each sampled school,
and each school will stay in the sample for a specific number of years.
In the motivating example, 10 wood pieces from each sampled
mill may have their mechanical strength measured.
For ease of presentation, we further simplify the notation
and let $\by_k = \{ \by_{k,i}: i \in s_k \}$ with $k = 0, 1, 2, \ldots, K$
where $s_k$ is the set of primary units in the sample on occasion $ k $.

The components within each $\by_{k,i}$ (for fixed $k, i$) are naturally dependent. 
This leads to within-population cluster/random effects.
When $k_1 \neq k_2$, $\by_{k_1,i}$ and $\by_{k_2,i}$
are data collected from the same PSU $i$ on two occasions. 
They are therefore naturally dependent, leading to longitudinal cluster/random effects.

Let $y$ be the response variable of a randomly selected 
SSU on occasion $k$.
We denote its population distribution by $F_k(y)$, for $k = 0, 1, \ldots, K$.
Since the population size is very large, for many types of response
variables, we may regard $F_k(y)$ as a continuous distribution.
This is also true when the values of $y$ in the finite 
population are random samples from a super-population
with continuous $F_k(y)$.
In some applications, $F_k(y)$ is naturally discrete and there
is no need to regard it as continuous.
In this case, our subsequent discussion remains applicable.
\vspace*{-2ex}

\subsection{Population and sampling plan}
Let $s_k$ be the indices of the PSUs
included in the $k$th sample ($k=0, 1, \ldots$).
To fix the idea, we highlight the following properties of the population and data 
from rotating sampling plans:

\begin{enumerate}
\item
The multiple samples are collected on several occasions
from the same grand population via a rotating sampling plan,
and the response values for the same unit may evolve.
\hspace*{-2ex}
\vspace*{-1ex}
\item
Each cluster $i$ forms a vector-valued time series in response 
$\by_{k, i}$ over $k=0, 1, \ldots, K$.
The time series formed by different clusters are mutually independent.
\hspace*{-2ex}
\vspace*{-1ex}
\item
The joint distribution $F_k$ of $\by_{k, i}$, which is common for all $i$,
is exchangeable with marginal distribution $ G_k $.
\hspace*{-2ex}
\vspace*{-1ex}
\item
The marginal distributions of 
any single response $G_k$, $k=0, 1, \ldots, K$, satisfy 
the DRM as specified in Equation (\ref{drm})
with a known basis function $\bq(\cdot)$.
For expository simplicity, the specific features of the DRM will be given later:
see \eqref{drm}.
\hspace*{-2ex}
\vspace*{-1ex}
\item
When $G_k = G_{k+1}$, the joint distributions of
$\{\by_{t, i}, t=0, 1, \ldots, k-1, k, k+1, \ldots, K\}$
and $\{\by_{t, i}, t=0, 1, \ldots, k-1, k+1, k, \ldots, K\}$
are identical for any $i$.
In other words, for fixed $i$, the distribution of 
$\{\by_{t, i}, t=0, 1, \ldots, k-1, k, k+1, \ldots, K\}$
is exchangeable for the $k$th and $(k+1)$th entries of the time series.

%\item
%The rotating plan leads to multiple samples ($k=1, 2, \ldots, K$)
%in the form of $\{\by_{k, i}, i=1, 2, \ldots, mN\}$ available for data analysis. 
\end{enumerate}

The properties above, except No.\ 4, are not too technical
and are plausible in the targeted applications.
The DRM assumption in No.\ 4 is also reasonable:
its validity mostly relies on the nonradical evolution of the population characteristics. 
Using this model leads to improved efficiency when it is approximately satisfied. 
The efficiency gain remains when this assumption is
mildly violated, as we will show in the simulation section.

We note that $G_k$ is the distribution of the
response value of a single SSU randomly selected from the
$k$th population. 
In this paper, we propose a permutation test for hypotheses
concerning functionals of $G_k$
based on multiple samples collected via the
rotating sampling plan described above.

\section{Permutation tests}
\label{permutation}
Let $F$ be the data-generating distribution
and $ R $ a test statistic designed to test a null hypothesis
against a specific alternative hypothesis: $H_0$ and $H_a$.
We assume that a larger $ R $ supports $F \in H_a$.
To construct a test of size $\alpha \in (0, 1)$, 
we look for a constant $ c = c_\alpha $ such that
\be
\label{rejectionRegion}
 \sup \{ P( R > c  | F): ~F \in H_0 \}= \alpha.
\ee
Let the observed value of $R$ be $r$.
The test rejects $H_0$ if $r > c_\alpha$.
One may equivalently compute a $p$-value
\be
\label{pvalue}
p = \sup \{ P( R > r  | F): ~F \in H_0 \}
\ee
and reject $H_0$ when $p \leq \alpha$, for that will imply 
$ r > c$ and hence rejection by the NP 
hypothesis-testing criterion.

In view of the above, the ultimate task of developing a valid test is to find 
an effective statistic $ R $ and a way to compute the resulting $p$-value while 
bypassing the need to specify $ c $ explicitly. 
In the context of tests based on multiple samples from a rotating sampling plan,
let
\[
R_n = R_n(\by_0, \by_1, \ldots, \by_K)
\]
be the test statistic of choice, with the subindex added to highlight
its dependence on the sample size.
Suppose the population distribution does not change
from occasion 0 to occasion 1: namely, $G_0 = G_1$. 
Then $(\by_{0i}, \by_{1i})$ and $(\by_{1i}, \by_{0i})$ have
the same distribution for all $i \in s_0 \cap s_1$. 
Taking advantage of this exchangeability,  we design
a permutation procedure as follows: 

\begin{description}
\item[Step I.]
For each $j \in s_0 \cap s_1$, generate a random permutation
$(a, b)$, independent of all other random variables, such that 
\be
\label{permute}
P\{(a, b) = (0, 1)\} = P\{(a, b) = (1, 0)\} = 0.5,
\ee
and let
$
(\by^*_{0, j}, \by^*_{1, j}) = (\by_{a, j}, \by_{b, j})
$.
Let $\by^*_{0, j}= \by_{0, j}$ and $\by^*_{1, j}= \by_{1, j}$
for $j \in s_0 - s_1$ and $ j \in s_1 - s_0$ respectively.

\item[Step I+.]
Permute $\{\by_{i}: i \in (s_0-s_1) \cup (s_1-s_0)\}$
to create clustered observations $\by^*_{0i}$ and $\by^*_{1i}$.

\item[Step II.]
Form a permuted multiple-sample
$\{ \by^*_{0, j}, j \in s_0\}$, $\{ \by^*_{1, j}, j \in s_1\}$,
and $\by^*_{ik} = \by_{ik}$ for $i \in s_k$ for $k=2, \ldots, K$.
\end{description}

We now present the proposed permutation test.

\vspace{1ex}
\noindent
{\bf Permutation test}.
{\it 
For each permuted multiple-sample, compute the value of the
test statistic 
\[
R_n^*= R_n(\by_0^*, \by_1^*, \ldots, \by_K^*).
\]
Generate permutation samples repeatedly and independently, 
say $M =10001$ times.
Compute the permutation test p-value 
\be
\label{pvalue}
p^* = \mbox{Proportion of~~ } \{R_n^* > r_n\}
\ee
where $r_n$ is the realization of $R_n$.
Reject the null hypothesis if $p^* < \alpha$ 
where $\alpha$ is the nominal level of the test.
}

\vspace{1ex}
\noindent
In applications, the practitioner conducts the test on a single
data set whereas in research projects analyses may be 
done with thousands of simulated data sets. 
Hence, it is computationally affordable to choose a large $ M $ in applications. 
The margin of error of $ p^* $ with the currently recommended $ M $  is about 
$(0.95*0.05)^{0.5}*1.96/M^{0.5} \leq 0.005$.
Allowing $ M $ to be an odd number helps to avoid minor operational issues.
In our simulation study, we use a much smaller $ M $
to allow for a large number of simulation repetitions.
Our reliance on the average performance of the tests, rather than
on accurate approximations in each repetition,  validates our choice
 of a smaller $ M $.

\vspace{1ex}
\begin{theorem}
\label{thm1}
Let $(\by_1^*, \by_2^*, \ldots, \by_K^*)$ be a permutation
multiple-sample obtained via Steps I and II above.
Assume that the null hypothesis $G_0 = G_1$ is true
and the model assumptions specified in the summary
subsection hold. Then we have the following results:

(a)
$R_n^* = R_n(\by_0^*, \by_1^*, \ldots, \by_K^*)$
has the same distribution as $R_n(\by_0, \by_1, \ldots, \by_K)$.

(b) 
Given $\{\by_0, \by_1, \ldots, \by_K\}$, $R_n^*$
has a discrete uniform distribution over all possible values
in the range of $ R_n(\by_0^*, \by_1^*, \ldots, \by_K^*) $.
\end{theorem}

\noindent
{\bf Proof}: 
(a)
When the null hypothesis holds, the joint distribution of 
$(\by_{0, i}, \by_{1, i}, \by_{2,i} \ldots, \by_{K, i})$ 
is the same as that of
$(\by_{1, i}, \by_{0, i}, \by_{2,i} \ldots, \by_{K,i})$
for all $i$ including all $i \in (s_0 - s_1)\cup (s_1 - s_0)$.
At the same time, $\by_{0, i}, \by_{1, i}, \by_{2,i} \ldots, \by_{K, i} $
with different $i$'s are mutually independent.
Therefore, the permutation Step I results in a new data set
whose joint distribution remains the same as
that of $\{\by_{k, i}, i \in s_k,  k=0, 1, \ldots, K\}$.
Therefore, $R_n^* = R_n(\by_0^*, \by_1^*, \ldots, \by_K^*)$
has the same distribution as $R_n(\by_0, \by_1, \ldots, \by_K)$.

(b)
The permutation prescribed in Equation~\eqref{permute} ensures
that every
permutation outcome has an equal probability. Hence,
$ R_n^* $ has a uniform distribution on these possible values.
This argument ignores rare but possible ties among these values. 
In such cases, we interpret the uniform distribution as 
a distribution with probabilities proportional to
the cardinality of each distinct permutation outcome.
\qed

\vspace{1.5ex}
\noindent
\begin{remark}\label{rem:1}
{\rm  The observed value $r_n$ of $R_n$ may be regarded
as one random outcome of $R_n^*$.
The conclusions in the above theorem hence ensure that
the type I error of the permutation test
equals the nominal level, excluding the round-off error.}
\end{remark}

\begin{remark}\label{rem:3}
{\rm  
The alternative hypothesis does not appear relevant in the proof or
theorem statement, but it matters for the actual test.
It determines the choice of the test statistic $R_n$.
We choose the $R_n$ that is the most sensitive to the
departure of the distribution in the direction of $H_a$, 
rather than arbitrary departures from the null hypothesis.
For this reason, the stochastic size of $R_n$ should increase when
the data-generating distribution $F$ is conceptually
deep in $H_a$ and far from $H_0$. 
In our target application, for example, if $H_a$
states that the population mean of $G_1$ is larger than that of
$G_0$, then an effective choice of $R_n$ is the difference in
the two sample means (namely $\bar{\by}_1 - \bar{\by}_0$). 
A larger difference in the population means
leads to a stochastically larger difference in the sample means.
%\textcolor{red}
{ If one chooses $R_n$ to be the difference in the two sample variances,
the resulting test may also suggest that $H_0$
(unequal mean) should be rejected, but for the wrong reason.}
}
\end{remark}

\begin{remark}\label{rem:5} 
{\rm Step I+ permutes the units in $(s_0 - s_1)\cup (s_1 - s_0)$.
The conclusion in Theorem \ref{thm1} breaks down when Step I+ is included:
namely, $R^*$ may have a slightly different distribution from $R_n$ under $H_0$.
However, under the null hypothesis, the difference introduced by this extra step 
is minor. At the same time, the units in $(s_0 - s_1)\cup (s_1 - s_0)$ contain 
crucial information when $H_a$ is true.
Hence, we recommend that Step I+ be included.
Our simulation study shows that the type I errors are not affected.}
\end{remark}

\begin{remark}\label{rem:6} 
{\rm In applications, things may not go as planned. A few primary units may
drop out from the rotating sampling plan.
A small modification is needed: permute
only units sampled on both occasions. }
\end{remark}

\section{Statistics of choice in permutation tests}
\label{statistics}
In this section, we propose some promising statistics
$R_n$ for the permutation test.
The choice of $R_n$ affects the statistical efficiency but not
the validity of the test.
\vspace*{-2ex}

\subsection{Straightforward choices of test statistics}
\label{sec4.1}
Let the null hypothesis be $H_0: G_0 = G_1$
and the alternative be $H_a: \xi(G_0) > \xi(G_1)$ 
with $\xi(G)$ being the mean, the quantile at some level of $G$,
or another population parameter. 

Two immediate choices are the classical $t$ and Wilcoxon rank-sum
statistics with the cluster structure in the data ignored:
\be
\label{t.test}
T = \frac{\bar {\by}_{1} - \bar {\by}_0}{\sqrt{ (1/n_0 + 1/n_1) s^2}}.
\ee
Here $\bar {\by}_{1}, \bar {\by}_0$ are the sample means, $s^2$ is
the pooled sample variance ignoring the cluster structure, and
\be
W = \sum_{i, j, u, v} \ind(y_{1, i, u} > y_{0, j, v})
\label{w.test}
\ee
where $\ind(\cdot)$ is the indicator function and
the summation is over all observations on occasions 0 and 1.
The Wilcoxon statistic is usually normalized in order to use the central limit theorem,
but this is unnecessary when the permutation approach is applied.

These two tests were originally designed to handle IID data. 
The $t$-test further requires that the data are from a normal distribution, 
and it detects the difference in the population means. 
The Wilcoxon rank-sum test is nonparametric and
primarily used to detect a location shift in two distributions,
although in theory it works only on the size of $P(X < Y)$.
Such limitations are often overlooked in applications,
and the tests serve general purposes surprisingly well.
However, this is not true for clustered data.  For such data, the tests
have inflated sizes (higher type I errors) if applied directly
without the proposed permutation procedure.
The generalization of the Wilcoxon test to independent clusters can
be found in \cite{datta2008signed}, \cite{datta2005rank},
and \cite{rosner2006wilcoxon}. Their results are not applicable
to clustered data with longitudinal random effects.

Let $\hat G_0$ and $\hat G_1$ be the distributions fitted
by any reasonable method. We may use
a straightforward statistic for the permutation test:
\be
\label{QorM}
R_n(\by_0, \by_1, \ldots, \by_K)
= 
\xi(\hat G_1) - \xi(\hat G_0).
\ee
Obvious choices for $\hat G_0$ and $\hat G_1$ are 
the empirical distributions ignoring the cluster structure
based on samples from $G_0$ and $G_1$.
Another possibility will be given in the next section.
We are most interested in this type of statistic for
population percentiles.
\vspace*{-2ex}

\subsection{DRM-assisted choices}       
\label{sec4.2}          
Under rotating sampling plans, the multiple-samples are
collected from closely related populations.
They naturally share some intrinsic latent structure.
Accounting for this structure leads to
more efficient estimates of $G_0$ and $G_1$
and therefore more powerful permutation tests.
We recommend the DRM introduced by
\cite{anderson1979multivariate}; we believe that it fits a broad range of situations. 
The DRM has been successfully used by many researchers,
including \cite{qin1997goodness}, \cite{qin1998inferences},
and \cite{keziou2008empirical}.

The DRM links the population distributions $G_k, k=0, 1, 2, \ldots, K$ by
\bas
\label{drm}
dG_k(y) = \exp\{ \btheta_k^\T \bq(y) \}  dG_0(y)
\eas
for some prespecified basis function $\bq(y)$ and parameter $\btheta_k$.
Note that $\btheta_0 =0$ when $G_0$ is chosen as the base distribution.
We require the first component of $\bq(y)$ to be $ 1 $ to make 
the first component of $\btheta$ a normalization parameter.
We use the EL of \cite{owen2001empirical}
as the platform for the inference.
In the spirit of the EL, we require $G_0$ to have the form
$
G_0 (y) = \sum_{k, i, u} p_{k, i, u} \ind(  y_{k, i, u}\leq y )
$.
We construct the {\it composite log likelihood function} 
\be
\label{logEL}
\ell^{\mbox{\tiny{C}}}_n(G_0, \ldots, G_K) 
=
\sum_{k, i, u} \log p_{k, i, u} + \sum_{k, i, u} \btheta_k^\tau \bq( y_{k, i, u})
\ee
with the summation over all possible indices $(k, i, u)$.
The DRM assumption implies the constraints
\be
\label{DRMconstraints}
\int \exp\{ \btheta_k \bq( y) \} d G_0 
= \sum_{k, i, u} p_{k, i, u} \exp\{ \btheta_k^\tau \bq( y_{k, i, u})\}
= 1
\ee
for all $k=0, 1, \ldots, K$. 
The log likelihood is  ``composite''
because the observations involved are dependent.
See \cite{lindsay1988composite} and 
\cite{varin2011overview} for an introduction to
and a general discussion of  the composite likelihood.

Given $\btheta_1, \ldots, \btheta_K$, maximizing
$\ell_n(G_0, \ldots, G_K)$ with respect to $G_0$ leads to the
profile log empirical likelihood function (in the same notation):
\ba
\ell^{\mbox{\tiny{C}}}_n(\btheta) 
&=&
\ell^{\mbox{\tiny{C}}}_n(\btheta_1, \ldots, \btheta_K) 
\nonumber \\
&=&
\sup \big \{
\ell^{\mbox{\tiny{C}}}_n(G_0, \ldots, G_K):
\sum_{k, i, u} p_{k, i, u} \exp\{ \btheta_j^\tau \bq( y_{k, i, u})\}
= 1; j=0, 1, \ldots, K
\big \}.
\label{DRMprofile}
\ea

Suppose $\hat {\btheta}_1, \hat{\btheta}_2, \ldots, \hat{\btheta}_K$ are
maximum EL estimators under DRM. The corresponding fitted
distribution functions are
%\textcolor{red}{
\be
\label{DRM.cdf}
\check G_j(y) 
= \sum_{k,i,u} \hat p_{k,i,u} \exp ( \hat{\btheta}^\tau_j \bq(y_{k,i,u}) )
\ind(y_{k,i,u} \leq y)
\ee
where 
$
\hat p_{k,i,u} = \big \{nr  \sum_{j=1}^K \exp(\hat{\btheta}^\tau_j \bq(y_{k,i,u})) \big \}^{-1}.
$
%}
The distribution function estimators can then be used in
\eqref{QorM} to form statistics for the permutation tests.
We give some specific statistics next.

\vspace{1em}
\noindent
{\bf Detecting changes in percentiles under DRM}

Let $\xi_\alpha(G)$ be the $(100\alpha)^{\rm th}$ percentile of $G$
with $H_0$ and $H_a$ being  $\xi_\alpha(G_0) =  \xi_\alpha(G_1)$ and
$\xi_\alpha(G_0) >  \xi_\alpha(G_1)$ respectively.
The solution for the two-sided alternative follows the same principle.

Under the DRM assumption, we give two choices.
The first choice is to let
\be
\label{ELQ}
R_n(\by_0, \by_1) = \xi_\alpha(\check G_1) -  \xi_\alpha(\check G_0)
\ee
where $\check G_0$ and $\check G_1$ are the fitted distribution
functions given in \eqref{DRM.cdf}.

The second choice is the empirical likelihood ratio statistic
with a computationally friendly alternation.
We first pool the samples from $G_0$ and $G_1$ to obtain the 
$100\alpha$th sample percentile:  $\hat \xi_\alpha$. 
We then compute the profile constrained composite  empirical likelihood
\ba
\label{constr}
\ell_n^{\mbox{\tiny{CC}}}(\btheta) 
&=&
\sup \big \{
 \sum_{k, i, u} \log p_{k, i, u} + \sum_{k, i, u} \btheta_k^\tau \bq( y_{k, i, u}):
 \nonumber \\
 &&  
 \sum_{k, i, u} p_{k, i, u} \exp \{ \btheta_s^\tau \bq(y_{k, i, u) }\} = 1
 \mbox{ for } s=0, 1, \ldots, K;
  \nonumber \\
&&
\sum_{k, i, u} p_{k, i, u} 
\exp \{ \btheta_s^\tau \bq(y_{k, i, u) }\}\ind(y_{k,i,u} \leq \hat \xi_\alpha) = \alpha
 \mbox{ for } s=0, 1
\big \}.
\ea
The recommended statistic for a permutation test is then
\[          
R_n 
= 
\sup \ell^{\mbox{\tiny{C}}}_n( \btheta) -  \sup \ell^{\mbox{\tiny{CC}}}_n(\btheta).
\]
We use an R-function called {\it RootSolve} to solve the
optimization problem. It solves equations formed by
the Lagrange multiplier method for constrained maximization.
With the corresponding derivative functions provided, this R-function works
well. The details are given in the Appendix.
\vspace*{-2ex}

\subsection{Populations satisfying model assumptions}
\label{sec4.3}
To support use of the proposed permutation test under the DRM with clustered data 
and as preparation for a meaningful simulation study, 
we consider the following examples.

\begin{example}\label{ex1}	
\label{normal}
{\bf Normal Data}.
{\rm  Let $\epsilon_{k, i, u}$, for $k=0, 1, \ldots$; $i = 1, 2, \ldots$;  
$u = 1, 2, \ldots$ be IID standard normal random variables.
Let $\eta_i, i=1, 2, \ldots$ be IID
standard normal random variables and 
$\eta_{k,i}, k=0,1, \ldots K, i=1, 2, \ldots$ another set of
IID standard normal random variables, where these are
mutually independent of $\epsilon_{k, i, u}$.
Let
\[
y_{k, i, u} 
= \mu_k + \sigma_{k,1} \eta_i + \sigma_{k,2}\eta_{k,i} + \sigma_{k,3}\epsilon_{k, i, u}
\]
for some nonrandom constants $\mu_k$ and $\sigma_{k,j}, j=1, 2, 3$.
}

{\rm Based on this construction, the random variables $y_{k, i, u}, u=1, 2, \ldots$ with fixed
$k, i$ are not independent but are identically and normally distributed.
Their joint distribution is exchangeable within the cluster indexed by $(k, i)$.
Furthermore, observations on the units in the same cluster taken on different occasions, 
e.g., $y_{k_1, i, u_1}$ and $y_{k_2, i, u_2}$,
 $ k_1\neq k_2$, are correlated
through the shared random effect $\eta_i$.
Given $ k $, the random variables
$y_{k, i, u}$ over $i=1, 2, \ldots$ and $u=1, \ldots, r$ have  identical marginal distributions.
We denote this distribution by $G_k$.
It is easy to verify that $G_0(y), G_1(y), \ldots, G_K(y)$ 
satisfy the DRM conditions with the basis function $\bq(y) = (1, y, y^2)^\tau$.
}

{\rm
In this model, $\mu_k$ is the nonrandom effect specific to the population
on occasion $k$.
The random effect $\eta_i$ is specific to the $i$th cluster
and shared over different occasions through the moderator $\sigma_{k,1}$.
The  random effect $\eta_{k, i}$ is specific to cluster $i$ and
independent over different occasions.
The response value of the $u$th unit in the $i$th cluster on occasion 
$k$ is given by $y_{k, i, u}$.

In summary, the longitudinal random effects are $\eta_i$ and the
cross-sectional random effects are $\eta_{k, i}$.
The marginal distributions satisfy the DRM with the basis function $\bq(y) = (1, y, y^2)^\tau$.
\qed
}
\end{example}

\begin{example}\label{ex2}	
{\bf  Gamma Data}.
{\rm 
A one-parameter Gamma distribution has
a degree of freedom parameter $\gamma$ with density function
\[
g^*(y; \gamma)  = y^{\gamma - 1} 
               \exp \{ - y\} \ind(y \geq 0)/\Gamma(\gamma) 
\]
where $\Gamma(\gamma)$ is the well-known Gamma function.
%We use $g^*(\cdot)$ rather than $g(\cdot)$ for gamma density to avoid
%notational conflict. 
}

{\rm 
Let $\bx$ be a vector and $a$ and $b$ two real numbers. 
We denote the vector comprised of $ax_i + b$ as $a \bx + b$.
With this convention, we create a complex cluster structure through the
operation for $k=0, 1, \ldots, K$ and $j=1, 2, \ldots$ :}
$\by_{k, j} = \lambda_k (\epsilon_j +  \epsilon_{k, j} + \bx_{k,j})$.
{\rm 
The elements of the stochastic models for $\by_{k, j}$ are specified as follows:}

\begin{enumerate}
\item
{\rm 
The $\epsilon_j$ are independent with distribution $g^*(y; \gamma_1)$.
Given cluster $j$, its value remains the same for all $k$, so  this
term leads to a longitudinal random effect.
}
\hspace*{-2ex}
\vspace*{-1ex}
\item
{\rm 
The $\epsilon_{k, j}$ are independent with distribution $g^*(y; \gamma_{2})$.
It is shared by the entries in cluster $j$ on occasion $k$, and this design leads to
the cross-sectional random effect.
}
\hspace*{-2ex}
\vspace*{-1ex}
\item
{\rm $\bx_{k,j}$ is a vector of independent
random variables with distribution $g^*(y; \eta_k)$
where $\eta_k$ is the degrees of freedom of occasion $k$.
They contribute most of the variations in the response vector $\by$.
The difference in $\eta_k$ leads to changes in the marginal distribution.}
\hspace*{-2ex}
\vspace*{-1ex}
\item
{\rm $\lambda_k$ introduces additional scale fluctuations over the occasions. 
}	
\end{enumerate}

{\rm The marginal distributions of $y_{k, i, u}$ are  $k$-specific
and denoted by $G_k$. 
%Its density will be denoted by $g_k$ as needed.
Because of the independence between $x_{k, j, u}$, $\epsilon_{k, i}$, and $\epsilon_j$, 
and  the property of the Gamma distribution, 
$G_k$ is also a Gamma distribution
with rate parameter $\lambda_k$ and degrees of freedom
$\gamma_1 + \gamma_{2} + \eta_k$.
Gamma distributions satisfy the DRM specified in
\eqref{drm} with $\bq(y) = (1, y, \log(y))^\tau$. }

{\rm In summary, by generating multiple samples from this model,
we obtain $\{\by_{k, i}, i \in s_k\}_{k=0}^K$ with
both longitudinal and cross-sectional random effects as described in
Section~\ref{setting}.  
%In addition, when $G_k = G_{k+1}$ for
%some $k$, exchanging $\by_{k,i}$ and $\by_{k+1,i}$ for any
%subset of $i$ in $s_k \cap s_{k+1}$
%does not change the joint distribution of the multiple samples.
\qed
}
\end{example}

\begin{example}\label{ex:3}
{\bf General data}.
{\rm Consider a population made of a large number of realized values
of a random sample from a super-population $F$. 
Denote these as $x_{k, i, u}$ with $(k, i, u)$ carrying no structural 
information at the moment. 
Let }
$
\by_{k, i} = \varphi(x_{k, i, 1}, \ldots, x_{k, i, r}; \epsilon_{k, i}, \epsilon_i)
$,
{\rm where }
\begin{enumerate}
\item
{\rm $\varphi(\cdot;  \epsilon_{k, i}, \epsilon_i)$ is an r-dimensional vector-valued function,
symmetric in $x_{k, i, 1}, \ldots, x_{k, i, r}$;}
\vspace*{-1ex}
\item
{\rm $\epsilon_i$: $i=1, 2, \ldots$ are IID; 
}
\vspace*{-1ex}
\item
{\rm $\epsilon_{k, i}$ are independent for different $(k, i)$, 
	and they are identically distributed given $k$. }
\end{enumerate}

{\rm In this general setting, the multiple samples $\{\by_{k, i}, i \in s_k\}_{k=0}^K$
have the cross-sectional and longitudinal random effects described in Section \ref{setting}.
In addition, when $G_k = G_{k+1}$ for
some $k$, exchanging $\by_{k,i}$ and $\by_{k+1,i}$ for any
subset of $i$ in $s_k \cap s_{k+1}$ 
does not change the joint distribution of the multiple sample.
At the same time, the population distributions clearly share some
general properties. A DRM with an appropriately
rich basis function $\bq(y)$, such as $\bq(y) = (1, y, y^2, \log y)$
when $y$ takes positive values, will be a good approximation
for the population distributions $G_0, G_1, \ldots, G_K$. }
\qed
\end{example}

\section{Simulation}
\label{simulation}
In this section, we present simulation results
to illustrate the effectiveness and necessity of the proposed permutation test.
We reveal the longitudinal random effects on the type I errors for
the t-test and Wilcoxon test (w-test).
\vspace*{-2ex}

\subsection{Data with normal distributions}
\label{simu.norm}
We generate data from the normal model as described in the last section.
The specific model parameters are chosen as follows:
\begin{enumerate}
\item
The number of occasions/populations is $K+1=5$.
\hspace*{-2ex}
\vspace*{-1ex}

\item
The number of units per cluster is either $r=5$ or $r=10$.
\hspace*{-2ex}
\vspace*{-1ex}

\item
The standard deviations are either 
$(\sigma_1, \sigma_2, \sigma_3) = (1, 1, 2)$ or $(1, 2, 3)$.
\hspace*{-2ex}
\vspace*{-1ex}

\item
The population means vector is one of: $(8, 8.4, \ldots)$, $(8, 8, \ldots)$, $(8, 7.6, \ldots)$, 
and $(8, 7.2, \ldots)$
with unspecified means randomly generated on each repetition as $8 + 0.5 N(0, 1)$.
\hspace*{-2ex}
\vspace*{-1ex}

\item
The number of clusters (primary units) in each sample is either $n=36$ or $n=48$.
The rotating sample plan replaces $m=6$ clusters on each occasion.
\end{enumerate}

The above choices lead to $2 \times 2 \times 4 \times 2= 32$ distinct settings.
Compounded with the permutations, this leads to a computationally challenging analysis.
We must reduce the computational burden.
Since the overall sample size increases either with more clusters or with larger cluster sizes,
in the simulation we avoid the option of increasing both sample size and cluster size.
These settings cover a broad range of qualitatively different situations: 
\begin{itemize}
\item
With different values of $(\sigma_2, \sigma_3)$, we learn the
performance of these tests for both relatively weak and strong cross-sectional cluster effects.
\hspace*{-2ex}
\vspace*{-1ex}
\item 
We learn if DRM-based methods benefit from their ability to borrow strength compared with methods that
use only information in the samples from the populations of interest.
\hspace*{-2ex}
\vspace*{-1ex}
\item
With different cluster sizes or numbers of clusters,
we learn about the consistency of these tests.
That is, the power increases to 1 when the sample size goes to infinity.
\end{itemize}

We consider the problem of testing whether the second population
has a smaller mean/percentile than the first population.
When the population means vector is set to $(8, 8, \ldots)$, the first two
population distributions are identical. The rejection rate of a test in this case
reflects its size.
When the vector is $(8, 8.4, \ldots)$, the population
mean of the second year is higher. The tests should therefore have rejection rates that are lower
than the nominal level. To control the amount of computation,
this setting is done only for $r=5$ and $n=36$.
The rejection rate of any effective test should be higher when the population 
means vector  changes to $(8, 7.6, \ldots)$ or $(8, 7.2, \ldots)$, 
where the alternative hypothesis holds.

In the simulations, we set the nominal level to $5\%$,
the number of repetitions to $1000$,
and the number of permutations to $201$.
We recommend a much larger number of permutations in applications.
In the simulations, the rejection rates are averages of 1000 repetitions. 
The precision of the individual $p$-values has little impact
on the overall  performance of the permutation test.
\vspace*{-2ex}

\subsubsection{To permute or not to permute}
We first demonstrate that the classical t-test and w-test
ignoring cluster structure indeed have inflated type I errors,
and that their desirable sizes are restored with permutation. 
Here $H_0$ claims that the first two populations have equal means, and
the test is one-sided so $H_a$ claims that the second mean is smaller.
We compute the $p$-values of the t-test and the w-test
using the inapplicable asymptotic results that ignore the cluster structure,
as well as the permutation approach. Table \ref{norm1.a} gives
the simulation results; the rejection rates are in the
Non-Perm and Permutation columns.

\begin{table}[h]
\caption{Rejection rates of t- and w-tests with clustered normal data}
\label{norm1.a}
\vspace*{-2ex}
\begin{center}
{\small
\begin{tabular}{|{c}|*{4}{r}|*{4}{r}|} \hline
\multirow{3}{*}{($\mu_0, \mu_1$)}
& \multicolumn{4}{c|}{($\sigma_1, \sigma_2, \sigma_3$)  = (1, 1, 2)}
& \multicolumn{4}{c|}{($\sigma_1, \sigma_2, \sigma_3$) = (1, 2, 3)}\\
 \cline{2-9}
& t.test & w.test &  t.test & w.test 
& t.test & w.test &  t.test & w.test \\
& \multicolumn{2}{c}{Non-Perm} & \multicolumn{2}{c|}{Permutation} 
& \multicolumn{2}{c}{Non-Perm} & \multicolumn{2}{c|}{Permutation}  \\ \hline
& \multicolumn{8}{c|}{$K+1=5,\: r= 5,\: n=36$}\\
 \cline{2-9}
(8.0, 8.4)& 0.3 &0.6 & 0.1 & 0.1     &     4.4 &   4.8 &  1.7  & 1.6\\
(8.0, 8.0)& 9.0 &9.6 & 5.1 & 4.9     &   14.2 & 13.8 &  6.2  & 6.4 \\
(8.0, 7.6)&45.8 &44.3 &28.8 &28.9&   31.6 & 30.9 & 16.5 &16.3 \\
(8.0, 7.2)& 85.2 &84.2 &73.4 &72.7&  61.4 & 60.3 & 37.5 & 37.6\\
\hline
& \multicolumn{8}{c|}{$K+1=5,\: r= 10,\: n=36$}\\
 \cline{2-9}
(8.0, 8.0)& 18.6 & 17.4 & 5.1 & 5.3      &   21.8 & 20.9 &  4.4  & 4.0\\
(8.0, 7.6)& 64.3 & 62.4 & 39.8 & 39.4   &   46.6 & 45.7 &  18.8  & 18.8 \\
(8.0, 7.2)& 95.8 & 95.6 & 84.6 & 82.7   &  75.2 & 74.7 & 45.0 & 44.9\\
\hline
& \multicolumn{8}{c|}{$K+1=5,\: r= 5,\: n=48$} \\
 \cline{2-9}
(8.0, 8.0)&  9.9  &  8.8  &   4.5 &  5.0 & 13.3 &13 5 &5.3 &5.3\\
(8.0, 7.6)& 56.3 & 55.2 & 39.9 & 38.5&  38.2 &36.6 &20.2 &19.7\\
(8.0, 7.2)& 94.1 & 93.7&  87.4 & 86.0 & 69.7 &68.4 &48.2 &46.7\\
\hline
\end{tabular}
}
\end{center}
\end{table}

The setting with $(\mu_0, \mu_1) = (8.0, 8.0)$  
lies on the boundary of the null hypothesis.
When $r=5$ and $n=36$, these two tests have rejection rates
as high as $9.0$\%, $9.6$\%, $14.2$\%, and $13.8$\%, if the cluster
structure is ignored (Non-Perm).
These rates are much closer to the nominal 5\% when the
cluster structure is handled via the permutation paradigm. The worst case is a
null rejection rate of 6.4\%, which is still in the range of
the simulation error given the 1000 repetitions. 
Other entries for $r=5$ and $n=36$ show that when 
$(\mu_0, \mu_1) = (8.0, 8.4)$, which makes the setting an
interior point of $H_0$, the rejection rates are below
the nominal level for all the tests.
Likewise, all the rejection rates increase
when $(\mu_0, \mu_1) = (8.0, 7.6)$
and increase further when $(\mu_0, \mu_1) = (8.0, 7.2)$.
These additional results are as expected
and therefore give general support to the validity  of our
simulation experiments.

Likewise, the results for $r=10$, $n=36$ and for $r=4$, $n=48$
are in the expected range.
We get the same message that ignoring the cluster structure
leads to inflated type I errors for classical tests.
Our permutation procedure is an effective way to
handle the clustering induced by the rotating sampling plan.
\vspace*{-2ex}

\subsubsection{Percentiles}
Percentiles are of particular interest in many applications, but
neither the t-test nor the w-test are designed to detect their changes.
In this section, we examine permutation tests based on the following
statistics introduced earlier:
\begin{eqnarray}
R_{\mbox{\tiny \sc em}}
& = &
\xi_{\alpha}(\hat G_1) -  \xi_{\alpha}(\hat G_0); \\
R_{\mbox{\tiny \sc el}} 
& =& 
\xi_{\alpha}(\check G_1) -  \xi_{\alpha}(\check G_0);\\
R_{\mbox{\tiny \sc elr}}
&=& 
\sup \ell^{\mbox{\tiny{C}}}_n( \btheta) -  \sup \ell^{\mbox{\tiny{CC}}}_n(\btheta).
\end{eqnarray}
The first is based on the empirical distributions $\hat G_0$ and $\hat G_1$.
The second is based on the fitted distributions $\check G_1$ and $\check G_0$ under the DRM
with the basis function vector $\bq(x) = (1, x, x^2)^\tau$ since we know that
the marginal distributions are normal.
The third is the likelihood ratio statistic under the DRM. 
No corresponding asymptotic theory is available, but
the permutation-based methods do not rely on
asymptotic theory.
We denote these tests by  EM, EL, and ELR in Table~\ref{norm1.b}.

We consider two null hypotheses: the first two populations have the
same $ 5^{\rm th}$ percentile or the same  $ 50^{\rm th}$ percentile.
The alternative hypotheses claim that the second population 
has lower percentiles. 
The rejection rates of these tests are presented in Table \ref{norm1.b}.

Clearly, the permutation tests tightly control the sizes of these tests. 
When the first two populations are identical, the rejection rates
range from 4.1 to 6.1, with the vast majority between 4.5 and 5.5.
When the first population has larger percentiles than the second one
(first line of the first block), the
type I errors are  much smaller than the nominal level, as expected.
When the second population has lower percentiles, the powers are 
higher than the nominal level, as they must be. 
The power increases further when the difference gets larger, and it 
 increases when either the cluster size or the sample size increases.

A comparison of the results in the left and right halves of the table
shows that the sizes of the permutation tests are not affected by
the strength of the random effects. When the random effects are
strong, the data contain less information. Hence, the powers on
the right half of the table are generally lower. The powers increase
when the cluster size increases or the number of clusters increases.

We expected the DRM-based tests to have higher powers,
and this is clearly true. Both EL and ELR are better than EM.
The improvement is more apparent
when the overall sample size is large and for the hypothesis regarding
the $ 50^{\rm th} $ percentile.
The simulation results in this table show that the performance
of the ELR is about the same as that of EL.

Finally, it is clear that change in a lower percentile is harder to detect than
change in the median under a nonparametric model assumption.
This explains the power differences for testing the changes in the 
$ 5^{\rm th} $ and $ 50^{\rm th} $ percentiles.
The simulation results are consistent with this intuition, and they also serve
as a sanity check. One may also conclude from the results in Tables \ref{norm1.a} and \ref{norm1.b}
that detecting changes in the median ($ 50^{\rm th}$ percentile) is harder than
detecting those in the mean.

\begin{table}[h]
\caption{Rejection rate of tests for equal percentiles with clustered normal data}
\label{norm1.b}
\vspace*{-2ex}
\begin{center}
{\small
\begin{tabular}{|{c}|*{3}{r}|*{3}{r}|*{3}{r}|*{3}{r}|} \hline
& \multicolumn{6}{c|}{($\sigma_1, \sigma_2, \sigma_3$)  = (1, 1, 2)}
& \multicolumn{6}{c|}{($\sigma_1, \sigma_2, \sigma_3$) = (1, 2, 3)}\\
\cline{2-13}
&\multicolumn{3}{c|}{$ 5^{\rm th}$ percentile} & \multicolumn{3}{c|}{$ 50^{\rm th}$ percentile}
&\multicolumn{3}{c|}{$ 5^{\rm th}$ percentile} & \multicolumn{3}{c|}{$ 50^{\rm th}$ percentile}
\\
&EM & EL & ELR  & EM & EL & ELR &EM & EL & ELR  & EM & EL & ELR\\
\hline
& \multicolumn{12}{c|}{$K+1=5,\: r= 5,\: n=36$}\\
 \cline{2-13}
1&0.8 &0.6 &0.8&  0.3 &0.1 &0.1  & 1.7 &1.9 &2.2   &1.5 &1.8 &1.9\\
2&5.4 &4.5 &4.7  &4.5 &5.0 &5.2  & 5.6 & 5.2 & 5.0   &6.0  &  6.1 & 6.1\\
3&17.7&19.2 &19.1 &25.6& 29.0 & 29.5 & 12.0 & 11.7 & 11.4 &15.2 & 16.7 &16.3\\
4&41.8&49.5 &48.4 &65.9& 75.5 & 75.6 & 22.5 & 26.2 & 26.0 &33.3 & 37.3 &37.4\\
\hline 
& \multicolumn{12}{c|}{$K+1=5,\: r= 10,\: n=36$}\\
 \cline{2-13}
2& 5.7 & 5.6 & 5.9 & 5.2 & 5.4 & 5.4  &5.3 & 5.0 & 5.1 & 4.2 & 4.2 & 4.1\\
3&13.2&13.9  &13.8 &18.8 &18.8 &18.9 &  21.8 &25.3  &24.6 & 36.5  &39.9 & 39.7\\
4& 52.1&59.7 &58.2 &79.0 &85.0 &84.6 &  27.1 &30.5  &30.2 & 42.0  &44.9 & 44.9\\
\hline
& \multicolumn{12}{c|}{$K+1=5,\: r= 5,\: n=48$}\\
 \cline{2-13}
2&  5.0 &   4.2 &   4.2 &  4.1  &4.5   &  4.6 &  5.3  &5.1  &4.9   & 4.7  &4.9 & 5.2 \\
3&21.4 & 24.9 & 24.6 & 34.6 &39.8 & 39.9&  12.8  &14.4  &14.1&16.6 & 19.9&  19.4\\
4&49.3 & 60.9 & 59.4 & 79.7 &87.4 & 87.2&  25.5 & 31.5 &31.2 & 41.1 & 47.9 & 47.7\\
\hline
\end{tabular}
}
\end{center}
\end{table}
\vspace*{-2ex}

\subsection{Data with Gamma distributions}
\label{simu.gamma}

In this section, we generate data from the Gamma model with
the parameters chosen as follows:

\begin{enumerate}
\item
The degrees of freedom vector $(\eta_0, \eta_1, \ldots)= (8.0, 8.4, \ldots)$, 
$(8.0, 8.0, \ldots)$, $(8.0, 7.6, \ldots)$, or $(8.0, 7.2, \ldots)$
with unspecified entries randomly generated in each repetition from $8 + 0.5 N(0, 1)$.
\hspace*{-2ex}
\vspace*{-1ex}

\item
The degrees of freedom vector is $(\gamma_1, \gamma_2) = (2.0, 1.5)$ or $(2.0, 3.0)$.
\hspace*{-2ex}
\vspace*{-1ex}

\item
The scale parameter is $(1.0, 1.0, \ldots)$ with
unspecified entries randomly generated in each repetition as $1 + 0.2U$,
$U$ being a uniform[0, 1] random variable.
\hspace*{-2ex}
\vspace*{-1ex}
\end{enumerate}

Similar considerations apply to this case.
The above settings enable us to examine the performance of these tests
in a broad range of situations.
We use $\bq(x) = (1, \log x, x)$ under the DRM assumption.
The other specifications are the same as those in the section on normal data.
\vspace*{-2ex}

\subsubsection{To permute or not to permute}
We now mimic the simulation conducted with the normal data.
The null hypothesis is that the first two populations have the same mean,
and the alternative is that the second population has a smaller mean.
The qualitative findings from the results in Table~\ref{gamma1.a} generally mirror those in 
Table~\ref{norm1.a}.
We again see the inflated type I errors for the non-permutation
tests and well-controlled type I errors for the permutation tests.
The powers of the permutation tests increase with increased cluster size
or increased number of clusters. The difference between the left and right
sides of the table is not remarkable, and it is not easy to decide 
which side has stronger cluster effects. 

{\small
\begin{table}[h]
\caption{Rejection rates of t- and w-tests with clustered Gamma data}
\label{gamma1.a}
\vspace*{-2ex}
\begin{center}
{\small
\begin{tabular}{|{c}|*{4}{r}|*{4}{r}|} \hline
\multirow{3}{*}{($\eta_0, \eta_1$)}
& \multicolumn{4}{c|}{($\gamma_1, \gamma_2$)  = (2, 1.5)}
& \multicolumn{4}{c|}{($\gamma_1, \gamma_2$)  = (1.5, 2.0)}\\
 \cline{2-9}
& t.test & w.test &  t.test & w.test 
& t.test & w.test &  t.test & w.test \\
& \multicolumn{2}{c}{Non-Perm} & \multicolumn{2}{c|}{Permutation} 
& \multicolumn{2}{c}{Non-Perm} & \multicolumn{2}{c|}{Permutation}  \\ \hline
& \multicolumn{8}{c|}{$K+1=5,\: r= 5,\: n=36$}\\
 \cline{2-9}
(8.0, 8.4)& 1.1 &0.9 & 0.3 & 0.4       &    1.9 &   1.6 &  0.7  & 0.9\\
(8.0, 8.0)& 8.4 &8.5 & 3.9 & 4.2       &   7.7 & 7.3 &  3.2  & 3.3 \\
(8.0, 7.6)&30.6 &31.0 &17.9 &19.0  &   32.5 & 32.7 & 23.1 &22.5 \\
(8.0, 7.2)&66.2 &67.1 &48.8 &50.1  &   69.0 & 68.2 & 56.5 &56.2 \\
 \cline{2-9}
& \multicolumn{8}{c|}{$K+1=5,\: r= 10,\: n=36$}\\
 \cline{2-9}
(8.0, 8.0)& 16.2 & 15.3 & 4.9 & 5.3      &   16.4 & 16.4 &  4.5  & 4.9\\
(8.0, 7.6)& 48.2 & 48.4 & 22.5 & 23.6   &   50.3 & 50.2 &  26.9  & 26.7 \\
(8.0, 7.2)& 85.7 & 85.1 & 61.3 & 61.0   &   69.0 & 68.2 &  56.5  & 56.2 \\
 \cline{2-9}
& \multicolumn{8}{c|}{$K+1=5,\: r= 5,\: n=48$}\\
 \cline{2-9}
(8.0, 8.0)& 11.7 & 11.9 & 4.5 & 4.8      &   9.5 & 9.0 &  5.0  & 5.3\\
(8.0, 7.6)& 42.9 & 42.4 & 26.3 & 26.5   &   37.2 & 36.3 &  26.5  & 26.4 \\
(8.0, 7.2)& 77.2 & 77.1 & 64.1 & 63.0   &   79.0 & 78.4 &  56.5  & 56.2 \\
\hline
\end{tabular}
}
\end{center}
\end{table}
}
\vspace*{-2ex}

\subsubsection{Percentiles}
We now move to testing hypotheses specifying equal 
$ 5^{\rm th}$ or $ 50^{\rm th}$ percentiles for
the first two populations in the multiple samples.
The alternative hypotheses are one-sided: $\xi(G_0) > \xi(G_1)$.
The simulation results are in Table \ref{gamma1.b}.
We could almost repeat here our earlier comments for the 
simulation results based on normal data. 
The type I errors of these tests are well controlled.
In fact, they are slightly low when $r=5$ and $n=36$. 
They are closer to the nominal level as either the cluster size or the
number of clusters increases. 

In all cases, the powers of these tests increase when either the cluster size or the
number of clusters increases. The EL and ELR have very similar
powers. They improve markedly over the EM test that does not
use model information in the DRM assumption. 
The differences between the left and right halves do not give much
more information, and both support the general claims in this paper.
In summary, the simulation results are as expected.

\begin{table}[h]
\caption{Rejection rate of tests for equal percentiles with clustered Gamma data}
\label{gamma1.b}
\vspace*{-2ex}
\begin{center}
{\small
\begin{tabular}{|{c}|*{3}{r}|*{3}{r}|*{3}{r}|*{3}{r}|} \hline
& \multicolumn{6}{c|}{($d_2, d_3$)  = (2, 1.5)}
& \multicolumn{6}{c|}{($d_2, d_3$)  = (1.5, 2.0)}\\
\cline{2-13}
&\multicolumn{3}{c|}{$5^{\rm th}$ percentile} & \multicolumn{3}{c|}{$50^{\rm th}$ percentile}
&\multicolumn{3}{c|}{$5^{\rm th}$ percentile} & \multicolumn{3}{c|}{$50^{\rm th}$ percentile}
\\
&EM & EL & ELR  & EM & EL & ELR &EM & EL & ELR  & EM & EL & ELR\\
\hline
& \multicolumn{12}{c|}{$K+1=5,\: r= 5,\: n=36$}\\
 \cline{2-13}
1& 0.9 & 0.6 & 0.6 & 0.5 & 0.2  & 0.3  & 1.3 &0.9 & 1.0   &0.6 &0.8 &0.9\\
2& 4.2 & 4.8 & 4.5 & 4.5 & 3.6  & 3.9  & 4.8 & 4.3 & 4.2   &3.5  &  4.1 & 3.7\\ 
3&13.4&16.7&16.7 &16.2&19.0 &18.5 & 14.4 & 16.7 & 16.2 &18.1 & 22.8 &22.8\\
4&29.6&34.9&34.7 &42.8& 50.0  & 49.5 & 29.6 & 37.3 & 36.3 &46.8 & 56.6 &56.8\\
\hline 
& \multicolumn{12}{c|}{$K+1=5,\: r= 10,\: n=36$}\\
 \cline{2-13}
2& 4.5 & 5.3 & 5.4 & 5.7 & 5.2 & 5.4       &   6.7  & 5.7 & 5.7 & 5.4 & 4.8 & 4.8\\
3& 18.4&19.9  &19.6 &21.1 &24.4 &23.9 &  18.8 &21.9  &21.7 & 25.2  &26.9& 26.8\\
4& 39.5&46.6  &46.4 &55.7 &63.2 &63.4 &  43.3 &51.4 & 50.5  &61.6 & 68.9 & 69.0\\
\hline
& \multicolumn{12}{c|}{$K+1=5,\: r= 5,\: n=48$}\\
 \cline{2-13}
2&  4.1 &  5.5 &  5.3   & 3.8   &  5.1  & 5.2   &5.3 & 5.9  & 5.7   & 5.3  &5.1 & 5.1 \\
3&15.9 & 20.0 & 19.3 & 22.8 &27.3 & 27.2& 14.6  &19.8  &18.7 &23.2 & 27.3& 27.2 \\
4&38.6 & 47.9 & 46.8 & 56.0 &64.7 & 65.0&  35.8 & 46.1 &46.1 & 56.5 & 68.3 & 68.1\\
\hline
\end{tabular}
}\end{center}
\end{table}
\vspace*{-2ex}

%\subsection{Risk of model mis-specification in DRM}
%
%Although DRM is of a non-parametric nature, we still have to provide
%a basis function $\bq(x)$ when it is applied. We used $\bq(x) = (1, x, x^2)$
%when data are normally distributed and $\bq(x) = (1, x, \log x)$ when data are gamma
%distributed. In applications, if we find the data distributions are normal--like, but are unwilling
%to use normal model, then using DRM with $\bq(x) = (1, x, x^2)$ seems  justified.
%A similar comment is applicable to gamma data.
%
%One may ask about the consequences that obtain 
%if neither of these two basis functions fit
%(when we know the truth). 
%Incidentally, we mistakenly used $\bq(x) = (1, x, x^2)$ for gamma distributed data
%in one simulation and obtained all the results.
%To save space, we do not include the table but give a summary
%instead.
%Other than the reduced advantage of EL and ELR over
%EM, all other conclusions derived from previous simulation results
%remain. The permutation tests have well controlled type I errors
%and good powers. More importantly, EL and ELR 
%still have higher powers than the EM.
%\vspace*{-2ex}
%
%We next use simulation experiments to further support the
%advantage of permutation tests based on DRM assumption.

\subsection{Data from no-name distributions}
\label{simu.noname}

In many applications, historical data sets of the same nature are available.
This paper recommends the use of the DRM to extract latent information 
from multiple samples to enhance efficiency.
When applying the DRM we must choose a basis function.
In simulations, we usually generate data from classical distributions,
and so an appropriate basis function is readily available.
In a parallel research project on a data-adaptive choice
of the basis function, we have found that the performance is enhanced under
the DRM assumption with  $\bq(y) = (1, \log |y|, y, y^2)$.
We demonstrate this point in this section:
the permutation tests still have well-controlled type I errors,
and they gain some efficiency under the DRM assumption.

We generate clustered data with all the features under 
a rotational sampling plan. 
We ensure that the population distributions share some
latent features, but simple basis functions are not available.
Nevertheless, we complete the simulation as
in the last two sections for EM, EL, and ELR
with $\bq(y) = (1, \log |y|, y, y^2)$ when the DRM is assumed.
Specifically, the data are generated as follows:

\begin{enumerate}
\item
\label{StepI}
Form a finite population  ${\cal P} = \{x_1, x_2, \ldots \}$ 
having a considerable size,  based on data from a real-world application.
%\textcolor{red}{
%We consider the case where all observations are positive values
%on a ratio--scale.}
\hspace*{-2ex}
\vspace*{-1ex}

\item
\label{StepII}
Randomly generate $\epsilon_j$, $\epsilon_{k, j}$ from the standard 
uniform distribution. 
Let
$b_{k, j}(x) = \exp\{\sigma_1 \epsilon_j  +\sigma_{k,2} \epsilon_{k, j} \log x \}$
for some positive constants $\sigma_1$ and $\sigma_{k,2}$.

Sample $ r $ values, $(\lambda_{k, j, 1}, \ldots, \lambda_{k, j, r})$,
from a Gamma distribution with $ 20 $ degrees of freedom and scale parameter $ 0.05 $.
Randomly draw $ r $ values $ x $ from $ {\cal P} $ with probability proportional to
$b_{k, j}(x)$ 
to form a cluster 
$\by_{k, j} =  (\lambda_{k, j, 1}x_{k, j, 1}, \ldots, \lambda_{k, j, r}x_{k, j, r})$.
\hspace*{-2ex}
\vspace*{-1ex}

\item
\label{StepIII}
Form multiple samples from a rotational sampling plan as
\[
\{  \by_{0, j}:  1 \leq j \leq mN\}; ~
\{  \by_{1, j}:  1 + m \leq j \leq mN + m\};~
%&\{  \by_{2, j}:  1+ 2m \leq j \leq mN+2m \}\\
 \cdots \cdots.
\]
\end{enumerate}

Here are some explanations. 
Step~\ref{StepI} creates a grand population, and
Step~\ref{StepII} uses a biased sample technique
to mimic the evolution of the
population distribution over occasions $k=0, 1, \ldots, K$.
The random numbers $\epsilon_j$ and $\epsilon_{k,j}$ induce
longitudinal and cross-sectional random effects; we use
$\sigma_1$ and $\sigma_{k,2}$ to adjust the strength of these effects.  
The $\lambda$ values are introduced to avoid identical
observed values in multiple samples. Since these values have mean 1
and small variance, this does not change the
expected value from the case where $\lambda = 1$ and is 
not random.

A large value of $\sigma_{k, 2}$ leads to stochastically
large sampled values. Hence, the hypotheses where
the two populations have equal percentiles (at some level)
and so on can be formulated through their values.
In this section, we consider testing only for equal 
$5^{\rm th}$ or $50^{\rm th}$  percentiles.
The alternative hypotheses remain one-sided: 
$\xi(G_0) > \xi(G_1)$.

We simulated data with both $\sigma_1 = 2$ and $\sigma_1 = 4$ to examine
the influence of the strength of the longitudinal random effects.
We successively set $(\sigma_{0,2}, \sigma_{1,2}, \ldots ) = (6.0, 7.5, \ldots)$,  
$ (6.0, 6.0, \ldots)$, $ (6.0, 4.5, \ldots)$, and $ (6.0, 3.0, \ldots)$
with unspecified entries generated in each repetition from
$3+ 2U$, where $U$ is a uniform [0, 1] random variable.
These are labeled as 1, 2, 3, 4 in the first column of the table
reporting the results. 

We defined the base case to have $ K+1 = 5 $, cluster size $ r=5$, and number of
clusters $n = 36$. We then repeated the simulation with an increased
cluster size $r=10$ in one setting and with an increased number of
clusters $n=48$ in another setting. 

The finite population ${\cal P} $ in this simulation is formed from   
data collected by students running experiments in a lab located at 
FPInnovations, Vancouver \citep{cai2016university}.
It is made up of 825 observed values
of the MOR of a specific type of wood product. 
The sample mean of this data set is $6.57$ $(1000~psi) $ and the sample variance is 
$2.82$.

The rest of the simulation settings are the same as before.
We do not include tests for changes in the population means
because they do not involve the DRM assumption.
The simulation results are given in Table \ref{gen1.b}.

The results confirm the points
we intended to make, and they provide routine sanity checks.
First, inference with nontailored basis functions under
the DRM assumption remains solid. The type I error fluctuates around the nominal
level in a low range. The powers of all the tests increase with the cluster size and with
the number of clusters. When the data-generating distributions are
not on the boundary of the null hypothesis,
the type I errors are below the nominal level, as shown in
the first row of the table.
Moreover, even when the data are from distributions 
that do not fully conform to the DRM specifications, 
the inferences made under the DRM assumptions remain
valid; the power comparison to EM remains favorable, although not
decisively so. Thus, we recommend the use of the DRM without reservation.

\begin{table}[h]
\caption{Rejection rate of tests for equal percentiles with clustered no-name data}
\label{gen1.b}
\vspace*{-2ex}
\begin{center}
{\small
\begin{tabular}{|{c}|*{3}{r}|*{3}{r}|*{3}{r}|*{3}{r}|} \hline
& \multicolumn{6}{c|}{$\sigma_1$ = 2}
& \multicolumn{6}{c|}{$\sigma_1$  = 4}\\
\cline{2-13}
&\multicolumn{3}{c|}{$5^{\rm th}$ percentile} & \multicolumn{3}{c|}{$50^{\rm th}$ percentile}
&\multicolumn{3}{c|}{$5^{\rm th}$ percentile} & \multicolumn{3}{c|}{$50^{\rm th}$ percentile}
\\
&EM & EL & ELR  & EM & EL & ELR &EM & EL & ELR  & EM & EL & ELR\\
\hline
& \multicolumn{12}{c|}{$K=5,\: r= 5,\: n=36$}\\
 \cline{2-13}
1& 1.4 &1.6 & 2.0 &  2.1 & 1.7 & 1.8  & 1.9 &2.8& 2.8& 2.3& 2.2& 2.2\\
2& 5.7 & 5.5 & 5.9 & 4.2 & 4.6 & 4.6  & 4.8 &5.0 & 5.2& 5.1& 4.8& 4.6\\ 
3& 8.3 & 9.2 & 9.4 & 11.2 & 10.8 & 10.8 & 9.0& 9.1& 9.0 & 11.2 &11.6 & 11.6\\
4&17.4& 18.2& 17.7& 24.2 &24.3& 24.6 & 16.9 &17.9& 17.9& 23.8& 24.8& 24.2\\
\hline 
& \multicolumn{12}{c|}{$K=5,\: r= 10,\: n=36$}\\
 \cline{2-13}
2& 4.8 & 5.4 & 5.6 & 4.4 & 4.6 & 4.4      &   6.2 & 5.6 &5.5 & 6.7 &6.4 & 6.6\\
3& 10.7& 11.4& 11.2 & 13.3 &13.0 &13.3 &  9.1 &8.8 & 8.6 & 9.4 & 9.9 & 9.9\\
4& 16.4 &17.1& 17.0 & 23.7& 25.2 &24.6 & 18.4& 18.9 &18.7 &22.7 &22.4 &22.6\\
\hline
& \multicolumn{12}{c|}{$K=5,\: r= 5,\: n=48$}\\
 \cline{2-13}
2&  4.9 & 5.2 & 5.4 & 5.0 &  5.7 & 5.9  &4.1 &5.2 &5.0 & 5.9& 5.3& 5.4 \\
3&11.8 &11.7 &11.4 &14.3 &14.6 &14.3 & 10.1 &10.1& 9.8 &12.0& 12.0& 12.3 \\
4&19.3 &20.9 &20.6 &29.4 &32.2 &31.7&  19.4 &20.1 &19.1& 27.6 &29.0 & 28.5\\
\hline
\end{tabular}
}
\end{center}
\end{table}
%
%\section{Summary and discussion}
%This paper develops novel permutation tests for 
%comparing/monitoring changes in mean, median or other parameters
%across multiple populations, in which the classical tests are found to have
%inflated type I errors. 
%The extensive simulation and theoretical analyses presented in this paper 
%lead us to conclude that the methods proposed in this paper are valid and 
%that they overcome deficiencies in classical tests 
%that are unable to accommodate the cluster effects seen in practice 
%when rotational panel designs are used. 
%These designs are important as they enable the benefits of the 
%classical paired t-test to be extended to much more complicated situations 
%in particular where concerns arise about temporal or spatial trends.  
%The software currently available along with the relatively 
%undemanding computation strategies, 
%will make the methods readily applicable by practitioners.  
%In particular the methods will be applicable in the lumber industries engaged 
%in monitoring for trends in their lumber's strength due to the 
%changing climate amongst other things. 
% 
 
{\bf Acknowledgements.}  
The research was funded by FPInnovations and a Collaborative Research 
and Development Grant from the Natural Sciences and Research Council 
of Canada. We are indebted to Mr.\ Conroy Lum.

\bibliographystyle{Chicago}
\bibliography{jiahua2019}

\newpage
\section*{Appendix: Computational issues}

The numerical implementation of most of our proposed permutation tests 
is straightforward.
The implementation of the ELR is conceptually simple but involves some tedious
steps. 
To compute $R_n$ defined following \eqref{constr}, we must solve the optimization problem 
$\sup_{\btheta} \ell_n^{\mbox{\tiny{CC}}}(\btheta)$.
The constraints in the definition of $\ell_n^{\mbox{\tiny{CC}}}(\btheta)$
can be rewritten as
\bas
\sum_{k, i, u} p_{k, i, u} 
[ \exp \{ \btheta_s^\tau \bq(y_{k, i, u})\} - 1] & =& 0, ~s=0, \ldots, K; \\
\sum_{k, i, u} p_{k, i, u} 
[ \exp \{ \btheta_s^\tau \bq(y_{k, i, u}) \}\ind(y_{k,i,u} \leq \hat \xi_\alpha) - \alpha ] & =& 0, ~s=0, 1.
\eas
Given $\btheta$, there always exists a ${\bf p}$, the vector formed by $\{p_{k, i, u}\}$,
that solves the above equation system provided vector ${\bf 0}$ is an interior
point of the convex hull of 
\bas
&
\{ \big ( [ \exp \{ \btheta_s^\tau \bq(y_{k, i, u})\} - 1]_{s=0}^K, 
[ \exp \{ \btheta_s^\tau \bq(y_{k, i, u}) \}\ind(y_{k,i,u} \leq \hat \xi_\alpha) - \alpha ]_{s=0}^1) \big ):\\
& 
\hspace{2em}
k=0, \ldots, K, i=1, \ldots, n; u=1, 2, \ldots, r \}.
\eas
The convex hull condition is universal and well-known in the literature of empirical likelihood
\citep{owen2001empirical}.
If this convex hull does not contain ${\bf 0}$, the adjusted empirical likelihood
approach of \cite{chenAsokanBovas2008} or the self-concordance empirical
likelihood of \cite{owen2013} may be used.
In the current application, we can show that as long as $\hat \xi_{\alpha}$ does not
fall outside either interval $( \min_{i, u} y_{s, i, u}, \max_{i, u} y_{s, i, u})$ for $s=0, 1$,
there always exist some $\btheta$ values such that the convex hull condition
is satisfied. 
In applications, if the joint sample percentile $ \hat \xi_\alpha$ 
has a nonextreme $\alpha$ value outside of 
one of these intervals, it is a strong indication that the population 
has significantly changed in some direction.
It is not urgent to look into such rare possibilities.

Once the existence of a solution is ensured, the optimization problem
can be solved via the Lagrange multiplier method. Because of the nice properties
of the DRM, we can find a simpler set of equations that can be solved by
the R-cran function RootSolve \citep{SoetaertHerman,RootSolve}. 
The details are as follows.

We first define a Lagrangian function:
\bas
g({\bf t}, \blambda, \btheta, {\bf p} )
&=&
\sum_{k, i, u} p_{k,i,u} +  \sum_{k,i,u} \btheta_k^\tau \bq(y_{k,i,u} ) \\
&&
- 
\sum_{s=0}^K t_s 
\Big [ 
\sum_{k, i, u} p_{k, i, u} \exp \{ \btheta_s^\tau \bq(y_{k, i, u) }\} - 1
\Big ]\\
&&
- \sum_{s=0}^1 
\lambda_s
\sum_{k, i, u} p_{k, i, u} 
\left [
\exp \{ \btheta_s^\tau \bq(y_{k, i, u) }\}\ind(y_{k,i,u} \leq \hat \xi_\alpha) - \alpha
\right ]
\eas
with ${\bf t}$ and $\blambda$ of length $K+1$ and two vectors of Lagrange multipliers.

The maximum of $ \ell_n^{\mbox{\tiny{CC}}}(\btheta)$ is attained
at the $\btheta$ value that solves
\[
\frac{g({\bf t}, \blambda, \btheta, {\bf p} )}{\partial {\bf t}} = 0;~~
\frac{g({\bf t}, \blambda, \btheta, {\bf p} )}{\partial {\blambda}} = 0;~~
\frac{g({\bf t}, \blambda, \btheta, {\bf p} )}{\partial {\btheta}} = 0;~~
\frac{g({\bf t}, \blambda, \btheta, {\bf p} )}{\partial {\bf p}} = 0
\]
together with some values of ${\bf t}$, $\blambda$, and $ {\bf p}$.

Some algebra shows that the solution in ${\bf t}$
is given by $t_s = nr $ where $nr$ is the total number of observations in each
population in the rotating sampling plan,
$s=0, 1, \ldots, K$, and the elements of ${\bf p}$ satisfy
\bas
p_{k,i,u} (\blambda, \btheta)
&=&
\left \{
(nr) \sum_{s=0}^K  \exp \{ \btheta_s^\tau \bq(y_{k, i, u} )\}
\right .
 \\
&&
\left .
 + nr(K+1)\sum_{s=0}^1 \lambda_s
 \left [
 \exp \{ \btheta_s^\tau \bq(y_{k, i, u) }\} \ind(y_{k,i,u} \leq \hat \xi_\alpha) - \alpha
 \right ]
 \right \}^{-1}.
 \eas
 Substituting the above expression into the Lagrangian equations, we obtain
 three sets of vector equations for $\btheta$ and $\blambda$:
 \bas
\sum_{k,i,u} p_{k,i,u} (\blambda, \btheta) \exp \{ \btheta_s^\tau \bq(y_{k, i, u} ) \} &=& 1; 
\\
\sum_{k,i,u}  p_{k,i,u} (\blambda, \btheta) \exp \{ \btheta_s^\tau \bq(y_{k, i, u} ) \}
  \{ \ind(y_{k,i,u} \leq \hat \xi_\alpha) - \alpha \}
 &=& 0; 
 \\
\sum_{k,i,u}  p_{k,i,u} (\blambda, \btheta) \bq(y_{k,i,u})  \exp \{ \btheta_s^\tau \bq(y_{k, i, u})\}
 [ 1 + \lambda_s \{ \ind(y_{k,i,u} \leq \hat \xi_\alpha) - \alpha\} ]
 & =& (nr)^{-1} \sum_{i, u} \bq (y_{k, i, u}),
 \eas
with $s=0, 1, \ldots, K$ for the first equation, $s=0, 1$ for the second equation,
and $s=0, 1, \ldots, K$ again for the third equation.
Note that the third equation takes vector values because $\bq(\cdot)$ is vector-valued.

Furthermore, the derivatives of the above equations (more precisely the related functions) 
with respect to $\btheta$ and $\blambda$ can be found without technical difficulties.
With this information provided to RootSolve, solving for
$\btheta$ in the simulation experiment was quite smooth. Of 1000 repetitions,
the R-function failed to find the solution about 10 times when hypothesis testing for 
the $5^{\rm th}$ percentile in the third example, and it succeeded in all the other cases. 
Because this is a low failure rate, we did not try to determine the exact cause and instead 
dropped these cases from the final tally. 
We did, however, increase the number of repetitions in the simulation so that the number of
successful repetitions in every setting was at least 1000.

\end{document}